\documentclass[
final            % use final for the camera ready runs
  ]
  {aipproc}

\usepackage[letterpaper]{geometry}
\layoutstyle{8x11double}
\begin{document}

\title{Development and results from a survey on students views of experiments in lab classes and research}


\author{Benjamin M. Zwickl}{
	address={Department of Physics, University of Colorado Boulder, Boulder, CO 80309}
}
\author{Takako Hirokawa}{
	address={Department of Physics, University of Colorado Boulder, Boulder, CO 80309}
}
\author{Noah Finkelstein}{
	address={Department of Physics, University of Colorado Boulder, Boulder, CO 80309}
}
\author{H. J. Lewandowski}{
	address={Department of Physics, University of Colorado Boulder, Boulder, CO 80309},
	altaddress={JILA, University of Colorado Boulder, Boulder, CO 80309}
}

\begin{abstract}
The Colorado Learning Attitudes about Science Survey for Experimental Physics (E-CLASS) was developed as a broadly applicable assessment tool for undergraduate physics lab courses.  At the beginning and end of the semester, the E-CLASS assesses students views about their strategies, habits of mind, and attitudes when doing experiments in lab classes. Students also reflect on how those same strategies, habits-of-mind, and attitudes are practiced by professional researchers.  Finally, at the end of the semester, students reflect on how their own course valued those practices in terms of earning a good grade.  In response to frequent calls to transform laboratory curricula to more closely align it with the skills and abilities needed for professional research, the E-CLASS is a tool to assess students' perceptions of the gap between classroom laboratory instruction  and professional research.  The E-CLASS has been validated and administered in all levels of undergraduate physics classes.  To aid in its use as a formative assessment tool, E-CLASS provides all participating instructors with a detailed feedback report.  Example figures and analysis from the report are presented to demonstrate the capabilities of the E-CLASS.  The E-CLASS is actively administered through an online interface and all interested instructors are invited to administer the E-CLASS their own classes and will be provided with a summary of results at the end of the semester.
\end{abstract}

\maketitle

%%%%%%%%%%%%%%%%%%%%%%%%%%%%%%%%%%%%%%%%%%%%
%% MAINMATTER
%%%%%%%%%%%%%%%%%%%%%%%%%%%%%%%%%%%%%%%%%%%%

\section{Introduction}

Laboratory courses offer an educational environment with immense flexibility and opportunity.  Within laboratory courses, students' theoretical ideas confront real world equipment and phenomena.  The laboratory course environment is a space where students can ask testable questions, design and carry out experiments, gather and analyze data, develop and refine models of observed phenomena, make testable predictions, develop scientific arguments, and present them to classmates and instructors.  The opportunities for such engagement is made possible through significant investments in dedicated laboratory space, apparatus, and small class sizes.  Despite the opportunities for engagement in many scientific practices around key ideas of the discipline,  there have been frequent concerns that such courses do not fulfill their potential.  In response, there are national calls to provide more authentic experiences with science to increase recruitment and retention in STEM \cite{PCAST2012}.  Also, over the last 15 years, a variety of curricula have emerged that use new educational technologies \cite{Thornton1990}, increase engagement with specific scientific practices (e.g., measurement and uncertainty \cite{Kung2005}), and use sophisticated apparatus close to contemporary physics research \cite{Galvez2005}.  Each of these curricula emphasize in various ways that students should develop habits of mind, experimental strategies, enthusiasm, and confidence around doing research---each curricula in its own way seeks to close the gap between the practices happening in the lab classroom and the practices of the research lab.  

The Colorado Learning Attitudes about Science Survey for Experimental Physics (E-CLASS) \cite{Zwickl2012}, has been developed as a formative assessment tool for measuring how students perceive the gap between doing physics experiments in their lab class and in professional research.  The E-CLASS is an \textit{epistemology} and \textit{expectations} (E\&E) survey \cite{Elby2011} about experimental physics, following in the footsteps of MPEX \cite{Elby2001}, VASS \cite{Halloun1998}, and CLASS \cite{Adams2006}.  Assessing students' \textit{epistemology} about experimental physics means students reflect on what counts as a valid experimental approach.  Because the term ``experiment'' is applied frequently in both school settings and in research settings, we assess students' epistemology through asking questions in pairs.    Assessing \textit{expectations} means assessing what students thought was important or valued in their course.   An E\&E survey is well-suited to current assessment needs in labs for two reasons. First, E\&E surveys are often widely applicable because they are not tied to specific course content.  Second, E\&E surveys are of most value when evaluating educational environments that have significant differences from professional practice.  It is in these classroom environments where we expect to see the largest splits in epistemology and expectation between students and experts.  As lab courses are transformed to align more closely to prepare students for research, we expect gaps between students and experts in epistemological beliefs about experiments will also narrow.
% [Say something about affect and confidence?]

\vspace*{-10pt}
\section{Design, Structure, and Validation}

Any formative assessment tool should accomplish three things.  First, it should measure something instructors care about. Second, it should measure something where we expect to see some variation.  And third, the results should be valid (i.e., believable).  We addressed the first goal by focusing the E-CLASS questions on aspects of laboratory instruction that reflect common goals of the laboratory curriculum.  We addressed the second goal by focusing on questions that address common criticisms of labs (e.g., students follow instructions without thinking).  And we addressed the third goal, demonstrating the validity of the results, through a series of 42 student interviews and by having 23 faculty and instructors respond to the E-CLASS statements as experts.

\begin{figure}
\includegraphics[width=0.47\textwidth, clip, trim=0mm 0mm 0mm 0mm]{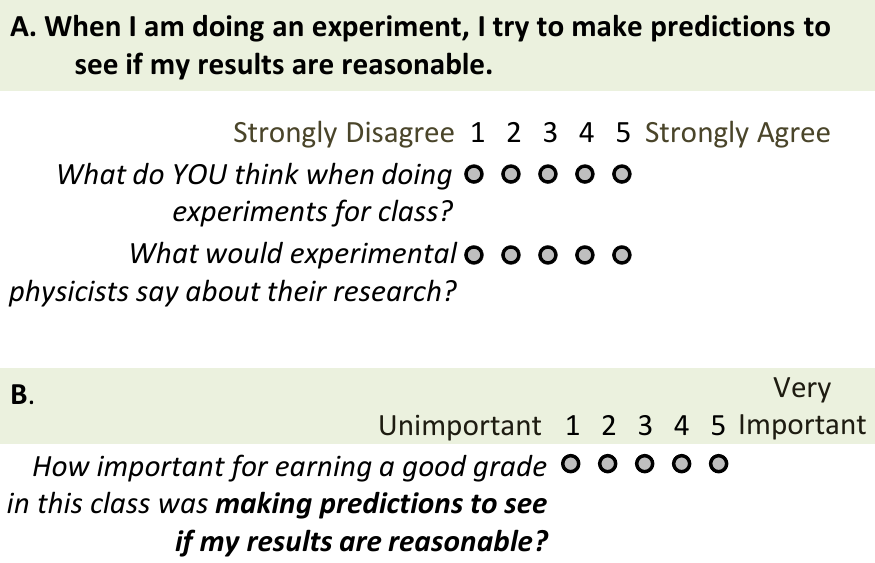}
\caption{Example questions from the E-CLASS.  (A) A pair of epistemology statements related to a single statement.  These questions are assessed pre and post-semester.  (B) The associated post-only question about students' reflections on their courses' expectations for earning a good grade.}
\label{fig:Example_questions}
\end{figure}

A typical subset of E-CLASS questions is shown in Fig.\ \ref{fig:Example_questions}.   The questions in Fig.\ \ref{fig:Example_questions} are all clustered around a single key idea of ``making predictions to see if my results are reasonable.''   The questions in Fig.\ \ref{fig:Example_questions}A are related to students' epistemology and attitudes about physics experiments in both classroom settings and in professional research.  The specific question wordings \textit{``What do YOU think when doing experiments for class?''} and \textit{``What would experimental physicists say about their research?''} were crafted through validation interviews with students in order to clarify the context of students' response (classroom vs research), and clarify whether they are answering what they personally think or what they should think (i.e., what would an expert say?).  The question \textit{``How important for earning a good grade...''} reveals students reflections on their course's expectations.  

\begin{figure}
\includegraphics[width=0.47\textwidth, clip, trim=0mm 0mm 0mm 0mm]{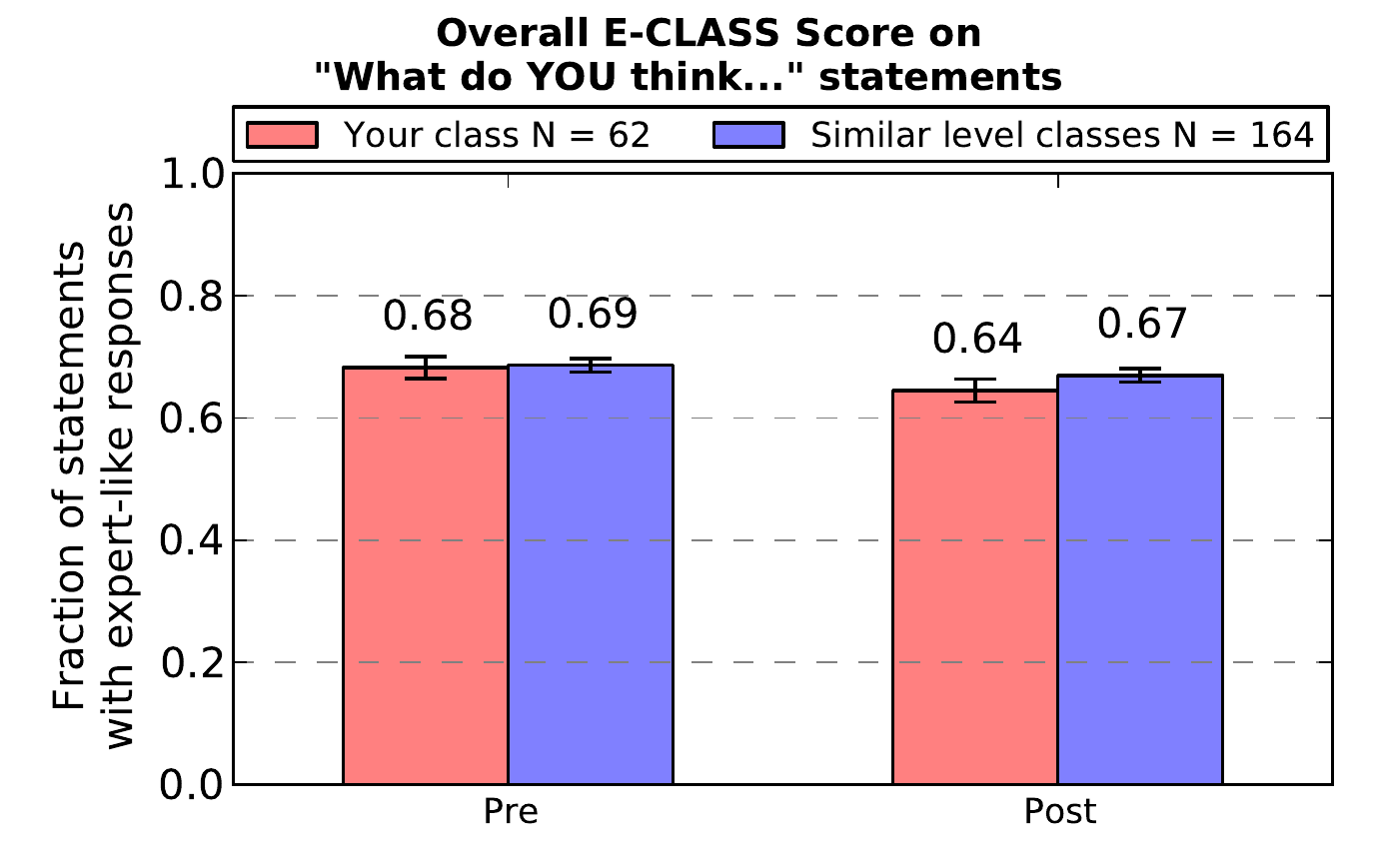}
\caption{Comparison between overall pre and post scores for students' personal views about \textit{``What do YOU think when doing experiments for class?''} \textbf{Your class} (Red) is compared with all students in \textbf{similar level classes} (i.e., introductory calculus-based physics classes) (Blue). The error bars show a 95\% confidence interval. The overall mean shown here averages over all students and all statements on the survey.}
\label{fig:Overall_ECLASS_score}
\end{figure} 

Although Fig.\ \ref{fig:Example_questions} shows just one key idea about doing physics experiments, there are altogether 23 different statements where students reflect on this trio of epistemology and expectations questions.   These statements were drawn from learning goals that have been established for both introductory \cite{Teachers1998} and upper-division physics labs \cite{Zwickl2013}.  In addition, there are 7 more statements about students' affect (e.g., ``If I wanted to, I think I could be good at doing research.'')  These seven affect statements are only assessed through the pair of epistemology questions \textit{``What do YOU think...''} and \textit{``What would experimental physicists say...?''}  

The need to distinguish between classroom and research contexts when asking questions about physics experiments was an important outcome of the student validation interviews.  Without specification, the term ``experiment'' evoked a variety of reasonable contexts, including in-class demonstrations, prior experiences in high school science labs, science fairs, undergraduate research experiences, and professional experiments like the Large Hadron Collider.  Clarifying the context and providing paired questions about classroom and research contexts was a key outcome of the validation interviews.  During the interviews, students were also asked to explain their interpretation of the 30 specific statements, and wording was revised to minimize technical jargon (e.g., ``doing error analysis'' was simplified to ``calculating uncertainties.'')  Some technical language highly relevant to upper-division physics labs was retained (e.g., systematic error). 

\begin{figure*}[h]
\includegraphics[width=0.70\textwidth, clip, trim=0mm 0mm 0mm 0mm]{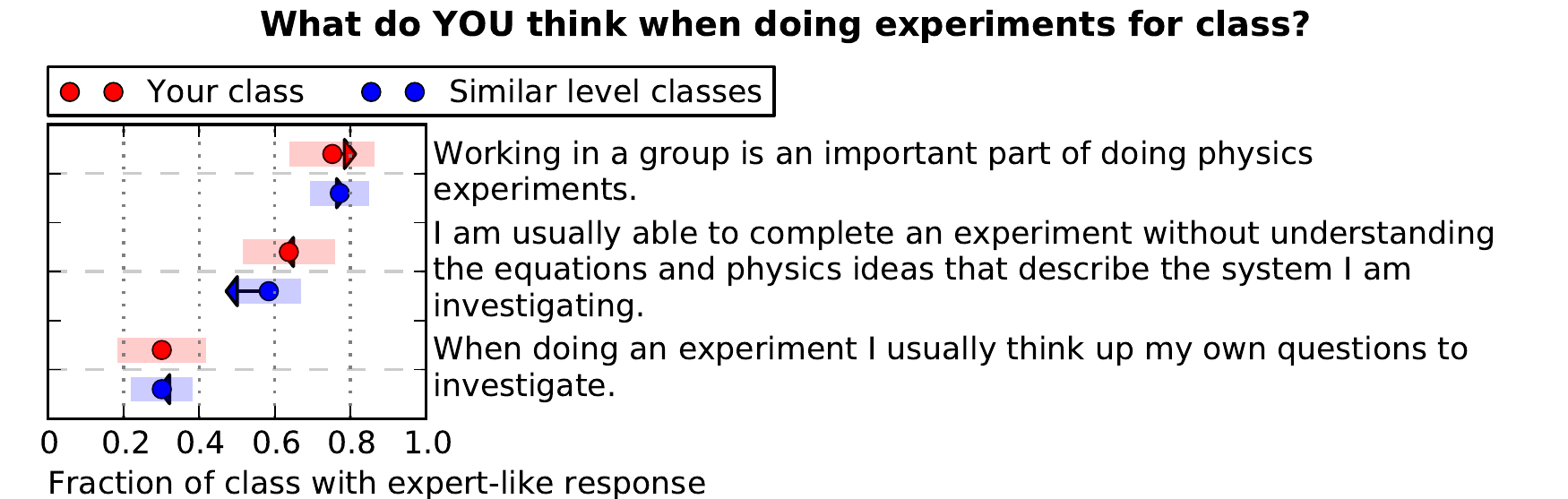}
\caption{Pre/Post changes in students' personal views about \textit{"What do YOU think when doing experiments for class?"} for \textbf{your class} (Red) and all students in \textbf{similar level classes} (i.e., introductory calculus-based physics classes) (Blue).  The circles show the pre-survey values. The arrows indicate the pre/post changes. The shaded bars are a 95\% confidence interval.}
\label{fig:Personal_pre_post_shifts}
\end{figure*}

It is also important to validate the survey with experts, which means demonstrating that experts have a consistent expert response to the answers.  We collected answers from 23 faculty and instructors from baccalaureate and masters granting institutions ($N$=7) and PhD-granting institutions ($N$=16).  Of the 30 statements, 24 had an expert consensus of 90-100\%, 3 had expert consensus of 80-90\%, while 3 had consensus of 70-80\%.  The three lowest statements were ``When I encounter difficulties in the lab, my first step is to ask an instructor.'' (18 disagreed, 5 neutral)  ``Working in a group is an important part of doing physics experiments.'' (17 Agree, 4 neutral, 2 disagree), and ``Nearly all students are capable of doing physics experiments.'' (16 agree, 4 neutral, 3 disagree).   For all statements, regardless of expert consensus, faculty can attend to or ignore statements depending on how they match their own course goals.

\vspace*{-10pt}
\section{Results}

We now show initial results from a typical course as they are given to instructors as a formative assessment tool. One of the highlights of the E-CLASS is that all instructors are presented with an explanation of their results, and a series of graphical plots of their class' data in comparison to other similar level classes (i.e., intro calculus-based labs, intro non-calculus-based labs, or upper-division labs).  

\begin{figure*}[h]
\includegraphics[width=0.70\textwidth, clip, trim=0mm 0mm 0mm 0mm]{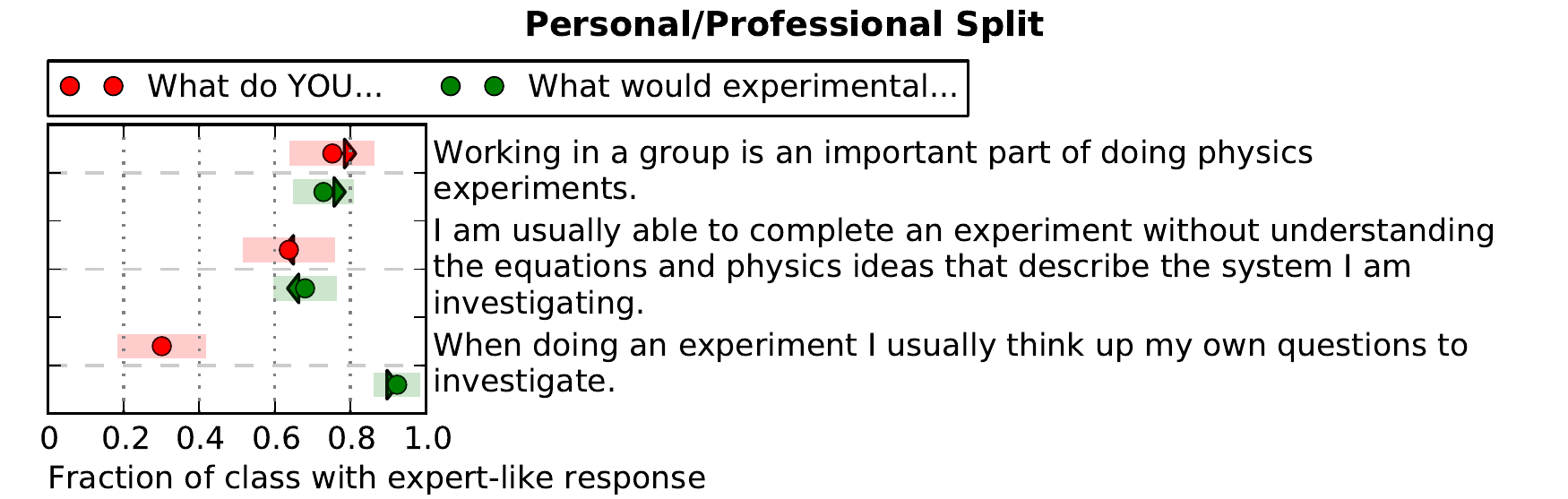}
\caption{Comparison of changes in students' personal views versus their views about professional physicists. Red shows the change in students' response to \textit{``What do YOU think when doing experiments for class?''}  Green shows the change in students' responses to \textit{``What would experimental physicists say about their research?''}  The circles show the pre-survey values. The arrows indicate the pre/post shift. The shaded bars are a 95\% confidence interval.}
\label{fig:Personal_professional_split}
\end{figure*}

\begin{figure*}[h]
\includegraphics[width=0.70\textwidth, clip, trim=0mm 0mm 0mm 0mm]{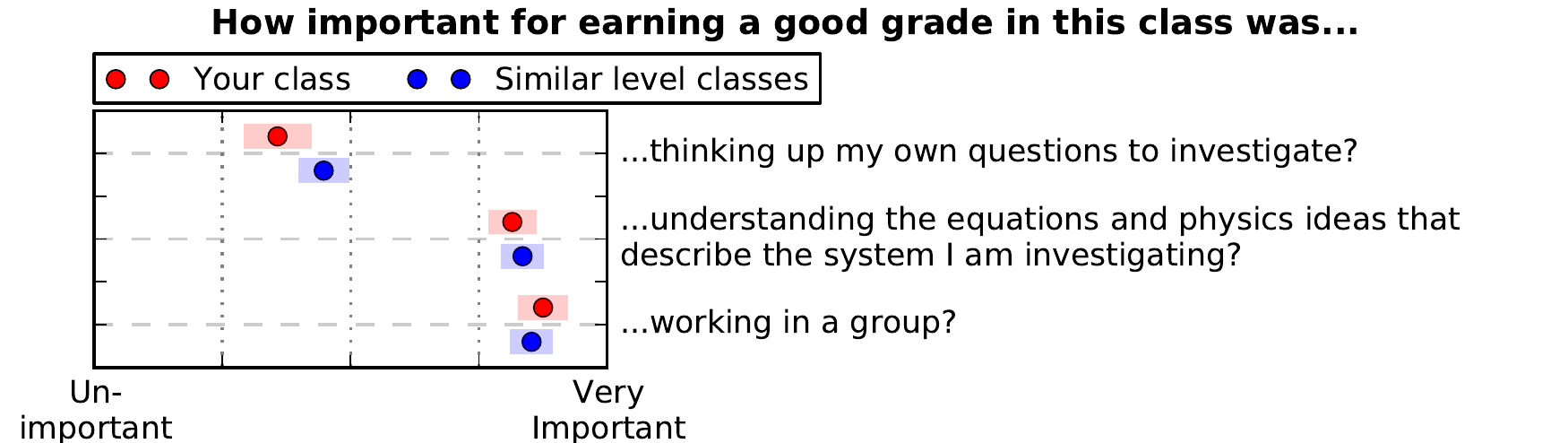}
\caption{Students' views of the importance of different activities for earning a good grade in \textbf{your class} (Red) and in \textbf{similar level classes} (i.e., introductory calculus-based physics classes) (Blue).}
\label{fig:Grade_plot}
\end{figure*}

Fig.\ \ref{fig:Overall_ECLASS_score} shows the overall E-CLASS pre and post scores for a calculus-based introductory lab at a large public PhD-granting university.  
This overall score represents the fraction of students' responses that are aligned with the expert consensus across all students in the class and all 30 questions in the survey.  While Fig.\ \ref{fig:Overall_ECLASS_score} is useful for a quick summary, the detailed view of individual statements reveals much more information about the class.  Those detailed views are shown in three forms (1) Students' personal views about \textit{``What do YOU think...?''} (see Fig.\ \ref{fig:Personal_pre_post_shifts}) 
(2) The split between students' views about doing physics experiments for class and experimental physicists doing research (See Fig.\ \ref{fig:Personal_professional_split}) and (3) Students' views about what was important for earning a good grade in the course (see Fig.\ \ref{fig:Grade_plot}).  
The plots shown here highlight statements with low expert-like agreement.  One thing that stands out from Fig.\ \ref{fig:Personal_pre_post_shifts} is that only about 30\% of students in this course and in similar-level courses agree that ``When doing an experiment, I usually think up my own questions to investigate.''  
While at the same time, Fig.\ \ref{fig:Personal_professional_split} shows that students do expect researchers to think up their own questions.  This is one of the statements with the largest epistemological splits between what students believe about classroom experiments versus research.  
Finally, Fig.\ \ref{fig:Grade_plot} shows that students in this class and similar-level classes rated ``thinking up my own questions to investigate.'' as relatively unimportant for earning a good grade.

By providing a measure of students' epistemology and their reflections on course grading expectations, we designed the E-CLASS survey to provide \textit{actionable} formative feedback for instructors.  For the key idea of ``thinking up my own questions to investigate,'' the E-CLASS suggests the course could integrate opportunities to a develop testable questions and investigate their answers.  Although not every instructor may pursue that goal, asking questions has been identified as a key scientific practice in the Next Generation Science Standards for K-12 \cite{NGSS2013}, in the AAPT goals for the introductory lab \cite{Teachers1998}, and is a key aspect of scientific research (100\% of the 23 experts agreed).   Clearly, the E-CLASS is measuring something of importance to the science education community, and our goal is that by providing an assessment with detailed reports we can encourage the long-term evaluation and refinement of laboratory curricula.

\section{Participation and Administration}
During Fall 2012 and Spring 2013, the online delivery of the E-CLASS has enabled us to administer the survey in 45 classes at 20 institutions in 3 countries, including an on-going evaluation of a Massive Open Online Course (MOOC).  The institutions represent a wide cross-section of institution types (7 Private, 13 Public), sizes (5 Small (1-3K), 3 Medium (3-10K), 12 Large(10K+)), and degree-granting statuses (1 associates, 5 baccalaureate, 3 masters, 11 PhD).  The 45 individual classes included 11 algebra-based introductory-level courses, 18 calculus-based introductory-level courses, and 16 laboratory courses beyond the intro-level, which were typically for physics and engineering physics majors.  The introductory courses tended to be larger, many in the range 50-200 students, while the upper-division courses were typically smaller, mostly in the range 8-25 students.  The median completion time on the Spring 2013 pre E-CLASS was 8 minutes ($N$=745), while for the post E-CLASS was 11 minutes ($N$=521).

Although we received responses from a large number of institutions and courses, the response rate in about half of those courses was disappointingly low.  Only 20 of 45 classes had a matching pre/post rate of greater than 40\%.  In some cases, this was due to lack of incentives for students (the lowest participation rates were in classes where faculty gave no credit for completing the survey),  while in other cases, it reflected an administrative obstacle of providing reminders to students at the appropriate times because semester start and end dates often differ by several weeks between institutions.  

We will continue to provide the E-CLASS as a formative assessment tool for any interested instructors provided they are willing to give their students credit or extra credit for completing the survey (even a very small amount is a good incentive).  In addition, we are actively seeking feedback from instructors on how they are using the E-CLASS as a formative assessment tool.  We want to refine both the survey and the reporting mechanisms to promote reflective teaching practices and encourage positive change in undergraduate laboratory curricula.   As data continues to be collected in a variety of laboratory courses in the United States and elsewhere, future papers will discuss those broader results.  

The authors would like to thank the CU Physics Department and PER Group for their support in developing the E-CLASS and the ALPhA community for widely administering the E-CLASS and providing feedback.  This work is supported by NSF TUES DUE-1043028.

%%%%%%%%%%%%%%%%%%%%%%%%%%%%%%%%%%%%%%%%%%%%%%%%
%% BACKMATTER
%%%%%%%%%%%%%%%%%%%%%%%%%%%%%%%%%%%%%%%%%%%%%%%%

%\begin{theacknowledgments}
%%We would like to thank ____. 
%This work was supported by National Science Foundation TUES award DUE-1043028.
%\end{theacknowledgments}



\bibliographystyle{aipproc}   % if natbib is available

\vspace*{-10pt}
%\bibliography{C:/SugarSync/Ben/Bibliographies/Publications-PRST_PER_ECLASS_Intro}
\bibliography{PERC2013}
\IfFileExists{\jobname.bbl}{}
 {\typeout{}
  \typeout{******************************************}
  \typeout{** Please run "bibtex \jobname" to optain}
  \typeout{** the bibliography and then re-run LaTeX}
  \typeout{** twice to fix the references!}
  \typeout{******************************************}
  \typeout{}
 }

\end{document}